\documentclass[aps,prl,reprint,twocolumn,nofootinbib,longbibliography]{revtex4-2}
\usepackage{amsmath,amssymb,graphicx,booktabs,array,needspace}
\usepackage{capt-of}
\usepackage[colorlinks=true,linkcolor=blue,citecolor=blue,urlcolor=blue,bookmarks=true,bookmarksnumbered=true,unicode=true]{hyperref}
\newcommand{\dd}{\mathrm d}
\begin{document}

\title{A Geometric No-Go Theorem for Regular Black Holes in Scalar-Tensor Theories}

\author{Shulang Peng}
\affiliation{College of Physics, Sichuan University, Chengdu 610065, China}

\author{Jun Tao}
\affiliation{College of Physics, Sichuan University, Chengdu 610065, China}

\author{Mian Zhu}
\affiliation{College of Physics, Sichuan University, Chengdu 610065, China}

\author{Yi-Fu Cai}
\affiliation{Department of Astronomy, School of Physical Sciences, University of Science and Technology of China, Hefei 230026, China}
\affiliation{CAS Key Laboratory for Research in Galaxies and Cosmology, School of Astronomy and Space Science, University of Science and Technology of China, Hefei 230026, China}

\begin{abstract}
  Perturbative instabilities generically puzzles singularity-free spacetimes, including non-singular bounce, traversible wormholes and regular black holes (RBHs). Can the scalar-tensor (ST) theories that stabilize the first two also stabilize RBHs? We show that the answer is no for static, spherically symmetric RBHs supported by a radial scalar field within a broad class of ST theories. In any RBHs with two-horizon structure, the ratio of the longitudinal scale to the sphere radius must attain an interior maximum. The proper-time derivative of that ratio is associated to  the squared propagation speed of odd-parity tensor perturbations by the background field equations in a broad class of scalar-tensor theories. A Lorentzian odd-tensor cone then requires a local minimum instead of a maximum. This conflicts indicate a geometric no-go theorem, covering a broad class of ST theories including spatially covariant gravity, Horndeski theory, quadratic GLPV, DHOST, and U-DHOST theories, and their admissible disformal extensions. Viable RBHs therefore require relaxing at least one geometric or dynamical assumption.
\end{abstract}
\maketitle
\footnotetext{Electronic addresses: zhumian@scu.edu.cn, yifucai@ustc.edu.cn}

\paragraph{Introduction.---}
The existence of spacetime singularities is a central problem in gravitational physics \cite{Hawking:1970zqf}. A major obstacle to constructing singularity-free geometries, including nonsingular cosmologies \cite{Novello:2008ra,Lehners:2008vx,Cai:2012va,Cai:2014bea,Brandenberger:2016vhg,Cai:2016hea}, traversable wormholes \cite{Morris:1988cz,Morris:1988tu}, and regular black holes (RBHs) \cite{1968qtr..conf...87B,Ayon-Beato:1998hmi,Hayward:2005gi}, is the persistence of perturbative instabilities \cite{Rubakov:2014jja}. Rigorous no-go theorems establish that such instabilities are generic for all three geometries \cite{Rubakov:2015gza,Rubakov:2016zah,Kobayashi:2016xpl,Libanov:2016kfc,Evseev:2017jek,Akama:2017jsa,DeFelice:2026qom} within Horndeski theory \cite{Horndeski:1974wa,Deffayet:2011gz}, the most general scalar-tensor (ST) theory with second-order equations of motion.

Going beyond Horndeski theory evades these theorems, enabling stable realizations of nonsingular cosmologies and static wormholes \cite{Cai:2016thi,Creminelli:2016zwa,Cai:2017tku,Cai:2017dyi,Cai:2017dxl,Cai:2017pga,Kolevatov:2017voe,Mironov:2018uou,Mironov:2019qjt,Ilyas:2020qja,Ilyas:2020zcb,Zhu:2021whu,Mironov:2022ffa,Volkova:2025uer,Ganz:2026mek}. Given the close mathematical parallel between the two geometries in perturbation theory \cite{Libanov:2016kfc,Rubakov:2016zah}, it is natural to ask whether the same theories also permit stable RBHs. Recent developments, however, point to a pessimistic answer: an explicit perturbative analysis identified stability obstructions across several RBH solution \cite{Boonserm:2026qwz} in Gleyzes-Langlois-Piazza-Vernizzi (GLPV) theory \cite{Gleyzes:2014dya,Gleyzes:2014qga}, a beyond-Horndeski ST theory known to accommodate stable bounces and wormholes \cite{Mironov:2019mye,Mironov:2022rmw}. Discovering the physical mechanism that sets RBHs apart from other singularity-free geometries and determining how broadly this obstruction holds across the ST landscape is therefore essential.

In this Letter, we identify this mechanism by extending the geometric argument of Ref.~\cite{Bronnikov:2001tv} (also see \cite{Easson:2026juf} for recent development). A standard nonextremal RBH solution contains an interior trapped region bounded by two horizons \cite{Hayward:2005gi}, which forces the ratio of the longitudinal scale to the sphere radius, $q$, to attain an interior maximum. We show that across a broad class of ST theories, which we name \emph{turning-point scalar-tensor (TPST) theories}, the background field equations relate the second proper-time derivative of $q$ at any turning point directly to the squared propagation speed of odd-parity tensor perturbations. Thus, a hyperbolic Lorentzian odd-tensor cone of tensor sector requires a local minimum instead of maximum, implying a confliction and thus a no-go theorem for RBHs in TPST theories. 

This geometric mechanism of the obstruction immediately explains why RBHs is different and extends the obstruction to broader scope. Because nonsingular cosmologies and wormholes lack the two-horizon trapped region, they are free from this geometric obstruction that puzzles RBHs. We also show that TPST class encompasses a broad class of ST theories, including spatially covariant gravity (SCG) \cite{Gao:2014soa,Gao:2014fra}, Horndeski theory, quadratic higher-order scalar-tensor (HOST) theories \cite{Langlois:2020xbc}, quadratic GLPV \cite{Gleyzes:2014dya,Gleyzes:2014qga}, degenerate higher-order scalar-tensor (DHOST) \cite{BenAchour:2016cay}, U-DHOST \cite{DeFelice:2018ewo,DeFelice:2021hps}, and theories related to them by admissible disformal transformations. Our result unifies the known exclusions \cite{Minamitsuji:2022vbi,DeFelice:2026qom,Boonserm:2026qwz,Babichev:2020qpr,Baake:2021jzv} into a single geometric no-go theorem for static, spherically symmetric RBHs supported solely by a real radial scalar field, excluding stables RBH solutions in TPST without requiring case-by-case investigations.

\begin{figure}[t]
\centering
\includegraphics[width=0.92\linewidth]{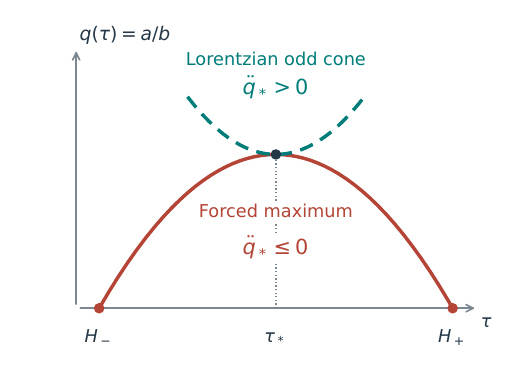}
\caption{The turning-point obstruction. Inside the $T$-region, the scale ratio $q(\tau)=a/b$ is positive and vanishes at the bounding horizons $H_\pm$, forcing an interior maximum (solid; $\ddot q_\ast \le 0$). A nondegenerate Lorentzian odd-parity tensor cone requires $c_{\rm odd}^{\,2}>0$ and hence a local minimum (dashed; $\ddot q_\ast>0$). The two requirements are mutually exclusive.}
\label{fig:turn}
\end{figure}

\paragraph{Geometric argument.---}
Consider a static, spherically symmetric metric \cite{1968qtr..conf...87B,Ayon-Beato:1998hmi,Hayward:2005gi}
\begin{equation}
ds^2=- A(r) \dd t^2 + \frac{\dd r^2}{B(r)} + r^2 \dd\Omega^2 ~.
\label{eq:metric}
\end{equation}
The regularity at the black hole center $r = 0$ requires $B(0) = 1$ \cite{Carballo-Rubio:2018pmi}, while asymptotic flatness enforces $\lim_{r \to \infty} B(r) = 1$. In a nonextremal geometry where all zeros of $B(r)$ are simple, $B(r)$ must therefore possess an even number of zeros. Since each zero corresponds to a horizon and a black hole contains at least one, a nonextremal RBH must have at least two horizons. We consider such canonical two-horizon RBHs with outer and inner horizons located at $r_+$ and $r_-$ ($0 < r_- < r_+$), respectively.

Around the nondegenerate outer horizon $r_+$, the metric functions expand as $A(r) = a_1(r-r_+) + \mathcal{O}((r-r_+)^2)$ and $B(r) = b_1(r-r_+) + \mathcal{O}((r-r_+)^2)$. A static Lorentzian exterior requires $a_1, b_1 > 0$, so $A(r)$ and $B(r)$ both become negative immediately inside $r_+$, and remain negative throughout the bounded interval $r \in (r_-, r_+)$ before vanishing again at the inner horizon $r_-$. This interior trapped domain is known as the $T$-region \cite{NovikovTregion}, in which $r$ is timelike and $t$ is spacelike, so the metric~\eqref{eq:metric} takes the homogeneous Kantowski--Sachs form
\begin{equation}
ds^2=-\dd\tau^2 + a(\tau)^2 \dd z^2 + b(\tau)^2 \dd\Omega^2 ~,
\label{eq:interior}
\end{equation}
with proper time $\tau \equiv \int \dd r/\sqrt{-B(r)}$, longitudinal coordinate $z = t$, longitudinal scale $a(\tau) \equiv \sqrt{-A(\tau)}$, and sphere radius $b(\tau) \equiv r(\tau) > 0$.

We define the ratio of the longitudinal scale to the sphere radius as
\begin{equation}
q(\tau) \equiv \frac{a(\tau)}{b(\tau)} = \frac{\sqrt{-A(\tau)}}{r(\tau)} ~,
\label{eq:ratios}
\end{equation}
which is strictly positive inside the $T$-region and vanishes at both horizons ($r \to r_\pm$). Consequently, $q(\tau)$ must attain at least one interior local maximum where $\dot q=0$ and $\ddot q \leq 0$, with dots denoting $\tau$ derivatives (Fig.~\ref{fig:turn}).

As a warm-up, consider the $k$-essence theory \cite{Armendariz-Picon:1999hyi,Garriga:1999vw}
\begin{equation}
S_{k}=\int \dd^4x\sqrt{-g}\left[\frac12 R + P(\phi,X)\right] ~,
\label{eq:kessence}
\end{equation}
where $X\equiv g^{\mu\nu}\nabla_\mu\phi\nabla_\nu\phi$ and we set $M_{\rm Pl}=1$. For a static radial scalar profile $\phi=\phi(r)$, inside the $T$-region the scalar field depends solely on proper time, $\phi=\phi(\tau)$, and its stress-energy tensor is spatially isotropic across the homogeneous constant-$\tau$ slices. Subtracting the longitudinal and angular Einstein equations therefore cancels $P(\phi,X)$ identically and yields $\frac{d}{d\tau} \left(ab^2 \frac{\dot q}{q} \right)= a$ \cite{Bronnikov:2001tv} (Appendix~\ref{sec:s1}), which in turn gives
\begin{equation}
\left. \frac{\ddot q}{q}\right|_{\dot q=0}=\frac{1}{b^2}>0 ~.
\label{eq:kessence_turn}
\end{equation}
At the stationary point $\dot q=0$, the positivity of $q$ forces $\ddot q>0$ and implying a local minimum, in direct contradiction with the interior maximum ($\ddot q\le 0$) demanded by the two-horizon geometry.

This geometric explanation clearly clarifies why beyond-Horndeski theories can stabilize nonsingular cosmologies and static wormholes \cite{Mironov:2019mye,Ilyas:2020qja,Mironov:2022rmw}, yet fail for two-horizon RBHs. First, bounces and wormholes lack a $T$-region, so their geometry does not force an interior maximum of $q$. Second, in Horndeski theory, bounces and wormholes are obstructed because a background factor equal to the tensor kinetic coefficient must cross zero (exceptions can exist within Horndeski theory but suffers from other challenges \cite{Ageeva:2020buc,Ageeva:2021yik,Akama:2022usl,GilChoi:2025hbs}). Beyond-Horndeski interactions lift this obstruction by decoupling the even-sector scalar-metric mixing combination from the tensor kinetic coefficient, allowing the mixing factor to cross zero while the tensor coefficients remain positive \cite{Mironov:2022ffa}. At the RBH turning point, by contrast, Eq.~\eqref{eq:turn} locks $\ddot q/q$ directly to the odd-parity tensor speed squared $c_{\rm odd}^{2}=F_{\rm T}/K_{\rm T}$, independent of even-sector mixing, see Appendix~\ref{sec:s6}).

\paragraph{TPST theory and the general no-go theorem.---}
We now show that the argument can be extended  to a broad class of ST theories, where the background field equations associate $(\ddot q/q)|_{\dot q=0}$ directly with the squared propagation speed $c_{\rm odd}^{\,2}$ of odd-parity tensor perturbations, so the tensorial instability is triggered by the same mechanism.

For a scalar field with a timelike gradient in the interior (that is, $\phi=\phi(\tau)$ with $\dot\phi\neq 0$), it is natural to adopt unitary gauge in the framework of spatially covariant gravity (SCG) \cite{Gao:2014soa,Gao:2014fra}:
\begin{equation}
S=\int \dd\tau\, \dd^3x\, N\sqrt h
\mathcal L(\tau, N, K,\mathcal S,\mathcal T,\mathcal R,\mathcal A) ~,
\label{eq:foliation}
\end{equation}
with $N$, $K_{ij}$, $\mathcal R$, $h_{ij}$ the lapse, the extrinsic and intrinsic curvature, and the induced spatial metric, respectively. We also include invariants such as $K=K^i{}_i$, $\sigma^i{}_j=K^i{}_j-K\delta^i{}_j/3$, $\mathcal S \equiv \operatorname{tr}\sigma^2$, $\mathcal T \equiv \operatorname{tr}\sigma^3$ and $\mathcal A \equiv D_i \ln N D^i \ln N$ in the Lagrangian (where $D_i$ is the covariant derivative compatible with $h_{ij}$), whereas higher derivatives of $K_{ij}$ and time derivatives of $N$ are absent.

The first key observation is that $q$ is governed by the anisotropic shear and is kinematically independent of the overall spatial volume $ab^2$. We thus adopt the parametrization
\begin{equation}
a(\tau)=a_0e^{\alpha(\tau)+2\beta(\tau)} ~,~ b(\tau)=b_0e^{\alpha(\tau)-\beta(\tau)} ~,
\label{eq:shapeparam}
\end{equation}
such that the spatial volume depends solely on $\alpha$ ($ab^2 = a_0 b_0^2 e^{3\alpha}$), and $q$ depends solely on the anisotropy parameter $\beta$ ($\dot q/q = 3\dot\beta$). A straightforward evaluation on the homogeneous background \eqref{eq:interior} gives  $K=3\dot\alpha$, $\mathcal S=6\dot\beta^2$, $\mathcal T=6\dot\beta^3$, $\mathcal R=2/b^2$, and $\mathcal A=0$. Varying the action \eqref{eq:foliation} with respect to $\beta$ then gives
\begin{equation}
\frac{\dd}{\dd\tau}\left[ab^2\left(12\mathcal L_{\mathcal S}\dot\beta
+18\mathcal L_{\mathcal T}\dot\beta^2\right)\right]
=2ab^2 \mathcal R \mathcal L_{\mathcal R} ~,
\label{eq:shapevolution}
\end{equation}
where subscripts on $\mathcal L$ denote partial derivatives, e.g., $\mathcal L_{\mathcal S}\equiv \partial\mathcal L/\partial\mathcal S$.

The next crucial step is to identify the local high-frequency dispersion relation of odd-parity tensor perturbations at the shear-free turning point $\dot q=0$:
\begin{equation}
-K_{\rm T}\omega^2+F_{\rm T}|\boldsymbol k|^2=0 ~,~
c_{\rm odd}^{\,2}=\frac{F_{\rm T}}{K_{\rm T}} ~.
\label{eq:oddchar}
\end{equation}
As shown in Appendix~\ref{sec:s1}, $K_{\rm T} \equiv \mathcal L_{\mathcal S}$ and $F_{\rm T} \equiv \mathcal L_{\mathcal R}$ at $\dot q=0$. Since $\ddot q/q = 3\ddot\beta$ when $\dot q = 0$, evaluating Eq.~\eqref{eq:shapevolution} at the turning point and substituting $\mathcal R=2/b^2$ together with Eq.~\eqref{eq:oddchar} yields our central turning-point identity:
\begin{equation}
\left.\frac{\ddot q}{q}\right|_{\dot q=0}
=\frac{F_{\rm T}}{K_{\rm T}b^2}
=\frac{c_{\rm odd}^{\,2}}{b^2} ~,
\label{eq:turn}
\end{equation}
which generalizes the $k$-essence result \eqref{eq:kessence_turn} ($K_{\rm T}=F_{\rm T}=1/2$ in $k$-essence theory). We note that neither Eq.~\eqref{eq:shapevolution} nor Eq.~\eqref{eq:oddchar} involves $\dot\phi$, so the identity~\eqref{eq:turn} can be analytically continued to where the scalar gradient momentarily vanishes.

We define a \emph{TPST theory} as any gravitational theory obeying the turning-point identity~\eqref{eq:turn}. At the interior maximum ($\ddot q\le0$) required by RBHs geometry, Eq.~\eqref{eq:turn} forces $c_{\rm odd}^{2}\le0$: the odd-parity sector must suffer either a gradient instability ($c_{\rm odd}^{2}<0$), an odd ghost ($K_{\rm T}<0<F_{\rm T}$), or a degenerate tensor cone ($c_{\rm odd}^{2}=0$, $K_{\rm T}=F_{\rm T}=0$) that signals strong coupling or loss of hyperbolicity \cite{DeFelice:2018ewo}. We thus arrive at our general no-go theorem: \emph{in any TPST theory, no static, spherically symmetric RBH supported solely by a real radial scalar field can simultaneously possess a bounded two-horizon $T$-region ($q\to 0$ at $r_\pm$) and a nondegenerate Lorentzian odd-parity tensor cone throughout the interior.}

\paragraph{Scope and extensions of TPST theory.---}
TPST covers a broad landscape of covariant ST theories widely studied in gravitational physics. Theories described by the foliated action \eqref{eq:foliation} clearly fall into this category. We first prove that the general quadratic HOST theory \cite{Langlois:2020xbc},
\begin{equation}
\scalebox{0.94}{$\displaystyle
S_{\rm quad}=\int \dd^4x\sqrt{-g} \left[F_0+Q\Box\phi+F_2R+\sum_{i=1}^{5}A_iL_i\right] ~,
$}
\label{eq:quad}
\end{equation}
belongs to TPST. To cover both standard areal coordinates ($J(r)=r$) and general radial coordinates $x$ where the sphere radius $J(x)$ in $ds^2 = -A(x) d t^2 + d x^2/B(x) + J(x)^2d \Omega^2$ may be nonmonotonic, we define $U \equiv A/J^2$ (so $U=-q^2<0$ inside the $T$-region and $U>0$ on the static exterior). Performing a volume-preserving shape variation ($\delta\ln\sqrt{|A|}=2\epsilon$, $\delta\ln J=-\epsilon$ at fixed $B$ and $\phi$, leaving $\sqrt{|A|}J^2$ invariant) leaves $X$, $\Box\phi$, and $L_{2,\dots,5}$ unchanged (Appendix~\ref{sec:s2}). Only $F_2R$ and $A_1L_1=A_1\phi_{\mu\nu}\phi^{\mu\nu}$ contribute, yielding the identity
\begin{equation}
\begin{gathered}
\Big[ \frac{J^4K_{\rm T}}{s}\,U' \Big]'=-2sF_2 ~,\\
K_{\rm T}=F_2-XA_1 ~,\qquad s=\sqrt{A/B} ~,
\end{gathered}
\label{eq:quadidentity}
\end{equation}
where primes denote $x$ derivatives. Inside the $T$-region, the odd-parity tensor perturbations are governed by the effective principal metric $F_2g^{\mu\nu}-A_1\phi^\mu\phi^\nu$ \cite{Takahashi:2019oxz}, corresponding to $F_{\rm T}=F_2$, $K_{\rm T}=F_2-XA_1$, and $c_{\rm odd}^{\,2}=F_2/K_{\rm T}$. Evaluating Eq.~\eqref{eq:quadidentity} at any turning point $\dot q=0$ ($U'=0$) and converting to proper time $\tau$ immediately reproduces the TPST identity~\eqref{eq:turn} with $b=J$. Crucially, this derivation uses no degeneracy relation among $F_2$ and $A_{1,\dots,5}$ and involves no division by $J'$ or $\phi'$: even if the scalar stalls at the turning point ($\phi'=0$, where unitary gauge fails), $X=B\phi'^2=0$ gives $K_{\rm T}=F_2$ and $c_{\rm odd}^{\,2}=1>0$, so the obstruction persists. 

Notably, quadratic HOST contains prominent subclasses, including Horndeski theory up to quartic order, quadratic GLPV, quadratic DHOST (such as classes Ia and IIa), and quadratic U-DHOST \cite{Boonserm:2026qwz,BenAchour:2016cay,Langlois:2020xbc,DeFelice:2018ewo,DeFelice:2021hps}, those theories therefore belong to TPST (an explicit U-DHOST example is given in Appendix~\ref{sec:s2}). We also show in Appendix~\ref{sec:s3} that the quintic Horndeski interaction $G_5(\phi,X)$, which is cubic in $\phi_{\mu\nu}$, likewise obeys the same identity \eqref{eq:quadidentity} with $(K_{\rm T},F_2)\to(\mathcal H,\mathcal F)$ and $\mathcal G=\mathcal H$ at $U'=0$, so the local odd-tensor speed squared $c_{\rm odd}^{\,2}=\mathcal F/\mathcal H$ again reproduces Eq.~\eqref{eq:turn}~\cite{Minamitsuji:2022vbi,Boonserm:2026qwz}. Therefore, the full Horndeski theory belongs to TPST class.

Moreover, the TPST no-go theorem extends to any theory $S=S_{\rm seed}\circ T$ connected to a TPST seed action $S_{\rm seed}$ by a field transformation $T \colon (g_{\mu\nu},\phi) \mapsto(\bar g_{\mu\nu},\phi)$ \cite{BenAchour:2016fzp,BenAchour:2020wiw,Takahashi:2022mew} that satisfies three local conditions near the interior maximum of $q$: (i) $T$ admits a smooth local functional inverse in field space so background solutions correspond; (ii) on the homogeneous $T$-region background, $T$ rescales the spatial scales isotropically ($\bar a=\sqrt C a$, $\bar b=\sqrt C b$) and the proper time via $d\bar\tau=\sqrt E d\tau$ with finite $C,E>0$, preserving $q$ and the sign of its turning-point acceleration,
\begin{equation}
\bar q=q ~,~
\left.\frac{d^2\bar q}{d\bar\tau^2}\right|_{\dot q=0}
=\frac1E \ddot q ~;
\label{eq:transportshape}
\end{equation}
and (iii) in the odd-parity sector, the linearized metric map is algebraic and nonsingular, $\delta\bar g_{\rm odd}=M \delta g_{\rm odd}$, so the odd-parity principal symbol transforms by congruence,
\begin{equation}
P_{\rm odd}(k)=M^{\mathrm T}\bar P_{\rm odd}(k)M ~,
\label{eq:congruence}
\end{equation}
preserving the signature and nondegeneracy of the odd-tensor cone. This immediately extends our no-go theorem to cubic GLPV and DHOST theories that are related to Horndeski by invertible disformal maps \cite{BenAchour:2016fzp,Takahashi:2022mew}, as well as admissible higher-derivative scalar-invariant and Weyl-curvature transformations \cite{Takahashi:2021ttd,Takahashi:2023vva}. Finally, imposing a mimetic constraint $\lambda[X-X_\star(\phi)]$ \cite{Arroja:2015wpa} also preserves both the shape identity and the odd principal symbol directly (Appendices~\ref{sec:s4} and \ref{sec:s5}; cf. Refs.~\cite{Franciolini:2018uyq,Lin:2017oow,Gao:2019twq}), thus those ST theories also belongs to TPST class.

In the end, we briefly discuss how our no-go theorem might extend to extremal and near-extremal RBHs. For a double-zero extremal horizon ($r_+=r_-=r_e>0$), one can expand the exterior metric functions around $\rho \equiv r-r_e > 0$ as $A(r)=a_2\rho^2+\mathcal O(\rho^3)$ and $B(r)=b_2\rho^2+\mathcal O(\rho^3)$ with $a_2,b_2>0$, and the near-horizon geometry approaches $\mathrm{AdS}_2\times S^2$ with $\mathrm{AdS}_2$ radius $L_2=b_2^{-1/2}$. For finite, nonzero horizon coefficients $K_{{\rm T},e}$ and $F_{2,e}$, the $\rho\to 0^+$ limit of Eq.~\eqref{eq:quadidentity} fixes the radial odd-tensor speed squared at the extremal horizon:
\begin{equation}
c_{r,\rm T}^{2}\big|_e=\frac{K_{{\rm T},e}}{F_{2,e}}=-\frac{L_2^2}{r_e^2}<0 ~.
\label{eq:extremal}
\end{equation}
By continuity, $c_{r,\rm T}^{2}<0$ persists on an open exterior interval outside $r_e$, where the odd-tensor principal metric acquires split signature $(2,2)$. On the static exterior where $\phi(r)$ is real ($\phi'^2=X/B>0$), the explicit class-Ia DHOST Hayward realization of Ref.~\cite{Babichev:2020qpr},
\begin{equation}
A(r)=B(r)=1-\frac{\mu\alpha r^2}{\beta r^3+\gamma} ~,~ X(r)=\frac{\alpha r^3}{\beta r^3+\gamma} ~,
\label{eq:hayward}
\end{equation}
has $c_{r,\rm T}^{\,2}=(6\beta X/\alpha-5)^{-1}\to -1$ at $r_e=(2\gamma/\beta)^{1/3}$ with $\alpha,\beta,\gamma>0$, agreeing with Eq.~\eqref{eq:extremal}. Varying the mass $\mu$ above $\mu_e = 3\beta r_e/(2\alpha)$ at fixed couplings, the exterior instability band $r_+<r<r_0\equiv(5\gamma/\beta)^{1/3}$ ($F_2<0<K_{\rm T}$) persists across the near-extremal window \cite{Baake:2021jzv})
\begin{equation}
1<\frac{\mu}{\mu_e}<2\left(\frac25\right)^{2/3} \simeq 1.1 ~,
\label{eq:masswindow}
\end{equation}
see Fig.~\ref{fig:hayward-exterior}. Thus the geometric argument can be extended to certain extremal RBH cases, see Appendix~\ref{sec:s7} for further examples.

\begin{figure}[t]
\centering
\includegraphics[width=0.9\linewidth]{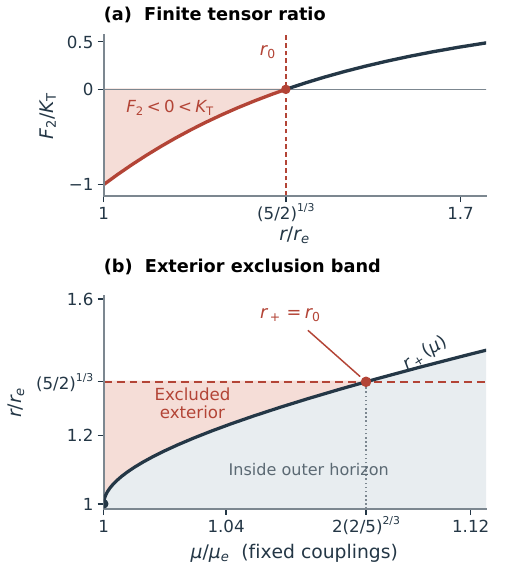}
\caption{Exterior exclusion in the class-Ia DHOST Hayward realization, Eq.~\eqref{eq:hayward}. (a) On the static exterior, the tensor ratio $F_2/K_{\rm T}$ is negative (shaded) for $1<r/r_e<(5/2)^{1/3}$, vanishing at $r_0=(5/2)^{1/3}r_e$. (b) The outer horizon $r_+(\mu)$ crosses the excluded band at fixed couplings when mass varies, leaving instable odd propagation ($F_2<0<K_{\rm T}$) outside $r_+$ throughout the near-extremal window $1<\mu/\mu_e<2(2/5)^{2/3}\approx1.1$.}
\label{fig:hayward-exterior}
\end{figure}

\paragraph{Conclusions.---}
\label{sec:conclusions}
We have demonstrated that the mechanisms that can stabilize nonsingular cosmologies and static wormholes doesn't necessarily stabilize RBHs. The reason is RBHs can have an additional stability constraint from its geometry that is absent in the first two geometries. In any RBH with a T-region, the vanishing of $q=a/b$ at both horizons forces an interior local maximum of $q$, whereas in any TPST theory the turning-point identity~\eqref{eq:turn} links $\ddot q/q$ at any stationary point to the odd-parity tensor speed squared $c_{\rm odd}^{\,2}$, requiring a local minimum for a hyperbolic Lorentzian tensor cone. This geometric no-go theorem encompasses well-known ST theory such as SCG~\eqref{eq:foliation}, Horndeski, general quadratic HOST \eqref{eq:quad} (including its quadratic GLPV, DHOST, and U-DHOST subclasses), and theories that are disformally connected to them.

Our result then suggests that constructing a viable RBH requires relaxing at least one of our geometric or dynamical assumptions: (i) allowing a time-dependent scalar profile $\phi=\phi(t,r)$ that does not inherit the static Killing symmetry or multiple scalar fields \cite{Nashed:2026tjp}; (ii) introducing additional anisotropic matter fields (such as nonlinear electrodynamics \cite{DeFelice:2024ops,DeFelice:2024seu,Ibadov:2026drq}) that modify the background shape equation; (iii) altering the interior causal structure so the trapped region is not bounded by two horizons where $q\to 0$ \cite{Lan:2026bdl,Bahamonde:2026iqb}; or (iv) incorporating higher-derivative operators beyond the TPST class \cite{Koshelev:2024wfk,Zhu:2025dur} that modify the relation \eqref{eq:turn}.

More broadly, our results highlight the value of exchanging ideas and methods between cosmology and BH physics. Together with recent developments such as quasinormal-mode collider spectroscopy \cite{Zhu:2026kka}, they illustrate how well-established methods in one area can help to address open problems in another and motivate the search for further connections across interdisplinary areas of gravitational physics.

\paragraph{Acknowledgments.---}
\label{sec:acknowledgments}
We are grateful to Shingo Akama, Yong Cai, Peng Wang, Yi Wang, and Shengfeng Yan for valuable comments. This work was supported in part by the National Key R\&D Program of China (2021YFC2203100). S. P. and M.Z. is supported by NSFC (12503005), Sichuan Science and Technology Program (2026NSFSC0804), and the Fundamental Research Funds for the Central Universities (YJ202551). J. T is supported by NSFC (12175212). Y.F.C. is supported in part by NSFC (12433002, W2611007), by CAS young interdisciplinary innovation team (JCTD-2022-20), by 111 Project (B23042), by CSC Innovation Talent Funds, by USTC Fellowship for International Cooperation, and by USTC Research Funds of the Double First-Class Initiative.

\appendix
\onecolumngrid
\setcounter{secnumdepth}{2}
\section{Brief summary of Appendices}
These appendices provide the detailed derivations, technical conditions, and explicit examples supporting the main text, following its logical progression:
Appendix \ref{sec:s6} contrasts regular black holes (RBHs) with nonsingular bouncing cosmologies and static wormholes, explaining the role of even-sector scalar--metric mixing;
Appendix~\ref{sec:s1} presents the geometric conditions for the interior maximum of $q$, the explicit $k$-essence warm-up derivation, and the background shape and odd-parity tensor perturbation analysis in spatially covariant gravity (SCG);
Appendix~\ref{sec:s2} derives the off-shell radial shape identity and odd-parity tensor principal symbol for general quadratic higher-order scalar--tensor (HOST) theories (including an explicit quadratic U-DHOST model);
Appendix~\ref{sec:s3} extends the radial shape identity to full Horndeski theory (specifically the quintic $G_5$ interaction);
Appendix~\ref{sec:s4} establishes the invariance of the turning-point obstruction under admissible disformal and higher-derivative field transformations;
Appendix~\ref{sec:s5} summarizes the covered theory families (Table~\ref{tab:coverage}) and proves direct extensions to higher-spatial-derivative SCG operators and noninvertible mimetic constraints;
and Appendix~\ref{sec:s7} proves the extremal and near-extremal exterior corollary and presents explicit class-Ia DHOST realizations.
Throughout, we adopt the metric signature $(-,+,+,+)$ with $X \equiv g^{\mu\nu}\nabla_\mu\phi\nabla_\nu\phi$.

\section{Comparison with Bounces and Wormholes}
\label{sec:s6}
\begin{figure}[htbp]
\centering
\includegraphics[width=0.7\linewidth]{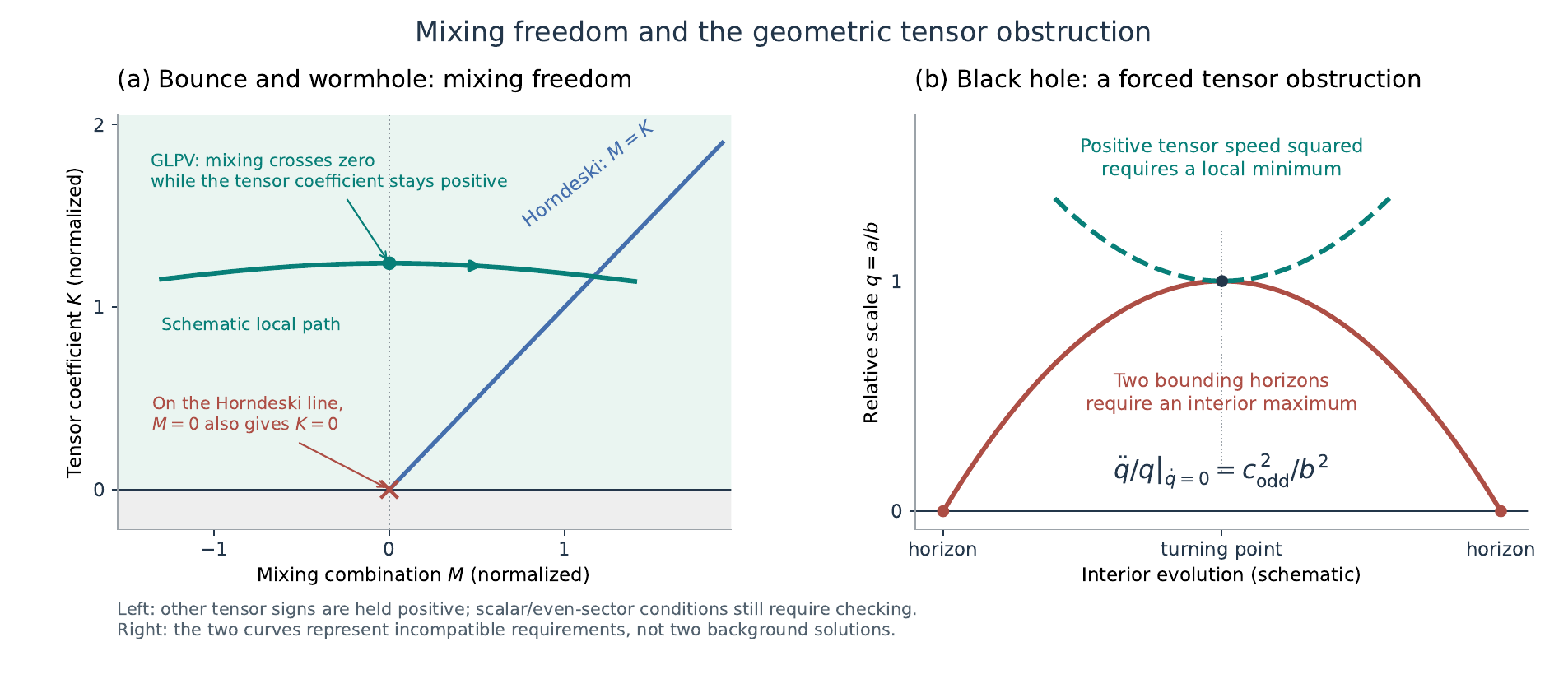}
\caption{Geometric incompatibility and the role of scalar--metric mixing. (a) In stable GLPV bounce and wormhole constructions, the even-sector mixing combination $M$ ($M_{\rm b}$ in cosmology, $M_{\rm WH}$ in wormholes) can cross zero independently of the positive tensor kinetic coefficient $K$ ($K=\mathcal G_{\rm T}$ or $K=\mathcal H$), whereas the Horndeski restriction $M=K$ forbids this crossing. (b) In a two-horizon RBH, the vanishing of $q=a/b$ at both horizons forces an interior maximum ($\ddot q_\ast \le 0$), whereas the turning-point identity~\eqref{eq:turn} directly locks $\ddot q/q$ to the odd-parity tensor speed squared $c_{\rm odd}^{\,2}$, requiring a local minimum ($\ddot q_\ast > 0$) independently of even-sector mixing.}
\label{fig:mixing}
\end{figure}

Here we provide the explicit formulas underlying our comparison between two-horizon RBHs and stable nonsingular cosmologies and static wormholes in beyond-Horndeski theories~\cite{Mironov:2019mye,Mironov:2022rmw,Mironov:2022ffa}, illustrated schematically in Fig.~\ref{fig:mixing}.

In spatially flat bouncing or Genesis cosmologies ($ds^2 = -\dd t^2 + a_{\rm c}(t)^2\dd\boldsymbol x^2$), the stability analysis of even-parity (scalar) and tensor perturbations is governed by the monotonic background relation~\cite{Mironov:2019mye}
\begin{equation}
\frac{\dd\xi}{\dd t} = a_{\rm c}(t)\bigl(\mathcal F_{\rm S} + \mathcal F_{\rm T}\bigr) > 0 ~, \qquad
\xi(t) \equiv \frac{a_{\rm c}\mathcal G_{\rm T} M_{\rm b}}{\Theta} ~, \qquad
M_{\rm b} \equiv \mathcal G_{\rm T} + \mathcal D\dot\phi ~,
\label{eq:s14}
\end{equation}
where $\mathcal G_{\rm T}, \mathcal F_{\rm T} > 0$ are the tensor kinetic and gradient coefficients, $\mathcal F_{\rm S} > 0$ is the scalar gradient coefficient, $\Theta$ is the standard denominator of the unitary-gauge EFT action, and $\mathcal D$ is the beyond-Horndeski (GLPV) coupling. Under standard asymptotic boundary conditions at $t \to \pm\infty$, the integral of $\dot\xi > 0$ forces $\xi(t)$ to cross zero at some finite time (unless $\Theta$ crosses zero, known as $\gamma$-crossing). In Horndeski theory ($\mathcal D = 0$), one has $M_{\rm b} = \mathcal G_{\rm T}$, so a zero of $\xi(t)$ forces the tensor kinetic coefficient $\mathcal G_{\rm T}$ to vanish, triggering an instability. In GLPV theory ($\mathcal D \neq 0$), however, the scalar--metric mixing factor $M_{\rm b} = \mathcal G_{\rm T} + \mathcal D\dot\phi$ is independent of $\mathcal G_{\rm T}$ and can cross zero smoothly while $\mathcal G_{\rm T}, \mathcal F_{\rm T}, \mathcal F_{\rm S} > 0$ remain strictly positive (Fig.~\ref{fig:mixing}(a)).

An analogous mechanism operates in static traversable wormholes~\cite{Mironov:2022rmw}, where the relevant factor is $M_{\rm WH} = \mathcal H - 2 F_4 B^2\phi'^4$, with $F_4$ the quartic beyond-Horndeski coupling. Setting $F_4 = 0$ (Horndeski) locks $M_{\rm WH} = \mathcal H > 0$, whereas $F_4 \neq 0$ allows $M_{\rm WH}$ to cross zero independently of $\mathcal H$. From the perspective of disformal transformations $\bar g_{\mu\nu} = g_{\mu\nu} + \Gamma(\phi,X)\phi_\mu\phi_\nu$ mapping GLPV to Horndeski~\cite{Mironov:2022ffa}, the kinetic Jacobian~\eqref{eq:ordinary-disformal} is proportional to $1 - X^2\Gamma_X = M_{\rm b}/\mathcal G_{\rm T}$; thus the zero-crossing $M_{\rm b} = 0$ is precisely the locus where the disformal map to Horndeski becomes singular even though the GLPV solution itself remains completely regular.

In a two-horizon RBH, by contrast, the obstruction arises in the pure odd-parity tensor sector rather than the even-parity scalar--metric sector: the turning-point identity~\eqref{eq:turn}, $(\ddot q/q)|_{\dot q=0} = F_{\rm T}/(K_{\rm T}b^2)$, involves only the odd-parity tensor coefficients $K_{\rm T}$ and $F_{\rm T}$ and contains no contribution from even-sector mixing couplings such as $\mathcal D$ or $F_4$ (Fig.~\ref{fig:mixing}(b)). Freedom to make an even-sector mixing combination cross zero therefore cannot alter the sign of $\ddot q$ at the turning point. The same logic explains why the DHOST evasion of the Horndeski bounce no-go \cite{Ilyas:2020qja}, which operates through the scalar dispersion relation, has no counterpart in the RBH case: the identity is saturated by tensor coefficients alone, and any operator that changes them changes both sides of Eq.~\eqref{eq:s1-undivided} simultaneously (Appendix~\ref{sec:s2}).

\section{Geometric Setup, k-Essence Warm-Up, and Spatially Covariant Gravity}
\label{sec:s1}

\subsection{Two-horizon \texorpdfstring{$T$}{T}-region geometry and the \texorpdfstring{$k$}{k}-essence warm-up}
Consider a static, spherically symmetric geometry on a connected radial interval $x \in (x_-, x_+)$ where the metric takes the form
\begin{equation}
ds^2 = -A(x)\,\dd t^2 + \frac{\dd x^2}{B(x)} + J(x)^2\,\dd\Omega^2 ~, \qquad A(x) < 0 ~, \quad B(x) < 0 ~, \quad J(x) > 0 ~,
\label{eq:s0-metric}
\end{equation}
supported by a real radial scalar field $\phi = \phi(x)$. In the standard areal-radius coordinate of Eq.~\eqref{eq:metric}, one simply has $x = r$ and $J(r) = r$; keeping $J(x)$ general allows us to treat coordinates in which the sphere radius is nonmonotonic on the same footing. Inside this trapped region ($T$-region), $x$ is timelike and $t = z$ is spacelike. Introducing the proper time $\dd\tau = \dd x/\sqrt{-B(x)}$ brings the line element to the homogeneous Kantowski--Sachs form of Eq.~\eqref{eq:interior}, with longitudinal scale factor $a(\tau) \equiv \sqrt{-A}$ and sphere radius $b(\tau) \equiv J > 0$.

The geometric input to our theorem is that the longitudinal-to-angular scale ratio $q \equiv a/b = \sqrt{-A}/J$ is strictly positive and twice continuously differentiable ($C^2$) on the open interval $(x_-, x_+)$, and extends continuously to $q(x_\pm) = 0$ at both bounding horizons $x_\pm$ (where $J(x_\pm)$ remains finite and positive). By the extreme-value theorem on the compact coordinate interval $[x_-, x_+]$, $q$ must attain a positive global maximum at some interior point $x_\ast \in (x_-, x_+)$. Because $\dd x/\dd\tau = \sqrt{-B} > 0$ is finite and nonzero at $x_\ast$, this interior maximum satisfies
\begin{equation}
\dot q(\tau_\ast) = 0 ~, \qquad \ddot q(\tau_\ast) \le 0 ~,
\label{eq:s0-max}
\end{equation}
where dots denote derivatives with respect to $\tau$. Notably, because compactness is used on the radial interval $[x_-, x_+]$ rather than on $\tau$, the existence of this interior maximum holds even if one or both horizons lie at infinite proper time (as occurs when the inner horizon is degenerate \cite{Hayward:2005gi,Frolov:2016pav,Carballo-Rubio:2022kad}), and requires only an adjacent pair of horizons bounding a single $T$-region regardless of the asymptotic geometry at infinity.

To see explicitly how this geometric maximum clashes with gravitational dynamics in the $k$-essence warm-up of Eq.~\eqref{eq:kessence}, $S_k = \int\dd^4x\sqrt{-g}\,[\frac{1}{2}R + P(\phi,X)]$, we compute the Einstein tensor for the Kantowski--Sachs metric~\eqref{eq:interior}:
\begin{equation}
G^z{}_z = -\frac{2\ddot b}{b} - \frac{\dot b^2}{b^2} - \frac{1}{b^2} ~, \qquad
G^\theta{}_\theta = G^\varphi{}_\varphi = -\frac{\ddot a}{a} - \frac{\ddot b}{b} - \frac{\dot a\dot b}{ab} ~.
\label{eq:s0-einstein}
\end{equation}
Because $\phi = \phi(\tau)$ depends only on proper time inside the $T$-region, its gradient is purely timelike ($\nabla_\mu\phi = -\dot\phi\,\delta_\mu^\tau$), so the $k$-essence stress-energy tensor $T_{\mu\nu} = P g_{\mu\nu} - 2P_X\nabla_\mu\phi\nabla_\nu\phi$ exerts isotropic spatial pressure across the longitudinal and angular directions: $T^z{}_z = T^\theta{}_\theta = P(\phi,X)$. Subtracting the longitudinal ($G^z{}_z = T^z{}_z$) and angular ($G^\theta{}_\theta = T^\theta{}_\theta$) Einstein equations therefore eliminates $P(\phi,X)$ identically:
\begin{equation}
0 = G^\theta{}_\theta - G^z{}_z = -\frac{1}{ab^2}\frac{\dd}{\dd\tau}\!\left[ab^2\left(\frac{\dot a}{a} - \frac{\dot b}{b}\right)\right] + \frac{1}{b^2}
\quad\Longrightarrow\quad
\frac{\dd}{\dd\tau}\!\left(ab^2\frac{\dot q}{q}\right) = a ~,
\label{eq:s0-kessence-diff}
\end{equation}
where we used $\dot q/q = \dot a/a - \dot b/b$. Evaluating Eq.~\eqref{eq:s0-kessence-diff} at any stationary point $\dot q = 0$ reproduces Eq.~\eqref{eq:kessence_turn}, $(\ddot q/q)|_{\dot q=0} = 1/b^2 > 0$, forcing a local minimum.

\subsection{Volume--shape variation in spatially covariant gravity}
We now derive the background shape equation~\eqref{eq:shapevolution} for the general SCG action~\eqref{eq:foliation}. In unitary gauge, a radial scalar field with a nonvanishing timelike interior gradient ($X = -\dot\phi^2 < 0$ near the turning point) defines the spatial foliation $\eta = \text{const}$. Retaining the homogeneous lapse $N(\eta)$ prior to variation (with $\dd\tau = N(\eta)\,\dd\eta$) and adopting the volume--shape parametrization $a = a_0 e^{\alpha+2\beta}$, $b = b_0 e^{\alpha-\beta}$ of Eq.~\eqref{eq:shapeparam}, the spatial volume factor is $v \equiv ab^2 = a_0 b_0^2 e^{3\alpha}$ and the ADM invariants take the form
\begin{equation}
K = \frac{3\alpha_\eta}{N} ~, \qquad \mathcal S = \frac{6\beta_\eta^2}{N^2} ~, \qquad
\mathcal T = \frac{6\beta_\eta^3}{N^3} ~, \qquad \mathcal R = \frac{2}{b^2} ~, \qquad \mathcal A = 0 ~,
\label{eq:s1-invariants}
\end{equation}
where the subscript $\eta$ denotes $\dd/\dd\eta$. Notice that separating $\alpha$ and $\beta$ decouples the volume $v(\alpha)$ from the shape parameter $\beta$ kinematically: holding $\alpha$, $N$, and $\phi$ fixed while varying $\beta$ gives $\delta v = \delta K = 0$ and $\partial\mathcal R/\partial\beta = 2\mathcal R$. Varying the reduced action $\int\dd\eta\,N v\,\mathcal L(\eta,N,K,\mathcal S,\mathcal T,\mathcal R,0)$ with respect to $\beta$ therefore yields
\begin{equation}
\frac{\dd}{\dd\eta}\!\left[v\left(12\mathcal L_{\mathcal S}\frac{\beta_\eta}{N}
+ 18\mathcal L_{\mathcal T}\frac{\beta_\eta^2}{N^2}\right)\right]
= 2 N v \mathcal R\,\mathcal L_{\mathcal R} ~.
\label{eq:s1-eta-shape}
\end{equation}
Setting $N = 1$ ($\eta = \tau$) after variation and writing $v = ab^2$ gives Eq.~\eqref{eq:shapevolution}. At any turning point of $q$ ($\dot q = 0$, so $\dot\beta = 0$), the cubic shear term and all time derivatives acting on $v$ or $\mathcal L_{\mathcal S}$ vanish because they multiply $\dot\beta$, reducing Eq.~\eqref{eq:s1-eta-shape} to $12\mathcal L_{\mathcal S}\ddot\beta = 2\mathcal R\,\mathcal L_{\mathcal R}$. Substituting $\ddot q/q = 3\ddot\beta$ and $\mathcal R = 2/b^2$ gives the undivided turning-point relation
\begin{equation}
\left.K_{\rm T}\frac{\ddot q}{q}\right|_{\dot q=0} = \frac{F_{\rm T}}{b^2} ~,
\qquad K_{\rm T} \equiv \mathcal L_{\mathcal S} ~, \quad F_{\rm T} \equiv \mathcal L_{\mathcal R} ~.
\label{eq:s1-undivided}
\end{equation}

\subsection{Odd-parity tensor perturbations and constraint reduction}
Next, we show that the very same background derivatives $K_{\rm T} = \mathcal L_{\mathcal S}$ and $F_{\rm T} = \mathcal L_{\mathcal R}$ evaluated at the shear-free turning point ($\dot\beta = 0$) govern the quadratic principal action for odd-parity tensor perturbations. At $\dot\beta = 0$, the background spatial slices are locally isotropic in the extrinsic curvature ($K_{ij} = \dot\alpha\,h_{ij}$, $\sigma_{ij} = 0$), so the linear odd-parity variations of all scalar arguments of $\mathcal L$ ($N, K, \mathcal S, \mathcal T, \mathcal R, \mathcal A$) vanish identically.

To extract the high-frequency principal symbol (corresponding to the odd-parity sector with angular multipoles $\ell \ge 2$), we work in a local orthonormal frame around the turning point where the $x^1$-axis lies along the longitudinal ($z$) direction and the $x^2, x^3$-axes lie tangent to the two-sphere $S^2$. Rotating the $S^2$ tangent axes so the spatial wave vector lies in the $x^1$--$x^2$ plane ($\partial_3 = 0$), the odd-parity metric perturbations are parameterized by the shift component $\delta N_3 = u_0(\tau, x^1, x^2)$ and the spatial metric components $\delta h_{13} = u_1(\tau, x^1, x^2)$ and $\delta h_{23} = u_2(\tau, x^1, x^2)$. Expanding the action~\eqref{eq:foliation} to quadratic order in perturbations and retaining terms with two derivatives of $u_a$ yields the principal Lagrangian
\begin{equation}
L_{\rm odd,pr}^{(2)} = \frac{\mathcal L_{\mathcal S}}{2}\left(f_{01}^2 + f_{02}^2\right) - \frac{\mathcal L_{\mathcal R}}{2} f_{12}^2 ~,
\qquad f_{ab} \equiv \partial_a u_b - \partial_b u_a ~.
\label{eq:s1}
\end{equation}
Here the quadratic variation of $\mathcal S = \operatorname{tr}\sigma^2$ provides the kinetic terms $f_{01}^2 + f_{02}^2$, while the quadratic variation of $\mathcal R$ provides the spatial gradient term $f_{12}^2$. By contrast, the cubic shear invariant $\mathcal T = \operatorname{tr}\sigma^3$ starts at third order in perturbations around $\sigma_{ij} = 0$, the expansion trace $K$ contains no odd-parity perturbations, and nonlinear functions of $\mathcal R$ produce no fourth-order derivative terms because $\delta\mathcal R = 0$ in the odd sector.

Passing to spatial Fourier space with wave vector $\boldsymbol k = (k_1, k_2, 0)$ and $k \equiv \sqrt{k_1^2 + k_2^2} \neq 0$, we decompose the spatial vector $(u_1, u_2)$ into longitudinal and transverse components:
\begin{equation}
u_i = \hat k_i u_L + \hat e_i u_T ~, \qquad \hat k_i = \frac{(k_1, k_2)}{k} ~, \quad \hat e_i = \frac{(-k_2, k_1)}{k} ~.
\end{equation}
Here the transverse mode $u_T$ is gauge invariant under odd-parity spatial diffeomorphisms ($x^3 \to x^3 + \xi^3$), whereas $u_L$ shifts by $ik\xi^3$. With $K_{\rm T} = \mathcal L_{\mathcal S}$ and $F_{\rm T} = \mathcal L_{\mathcal R}$, Eq.~\eqref{eq:s1} becomes
\begin{equation}
L_{\rm odd,pr}^{(2)} = \frac{K_{\rm T}}{2}\left(|\dot u_L - ik u_0|^2 + |\dot u_T|^2\right) - \frac{F_{\rm T}}{2} k^2 |u_T|^2 ~.
\label{eq:s2}
\end{equation}
For $K_{\rm T} \neq 0$, varying Eq.~\eqref{eq:s2} with respect to the nondynamical shift perturbation $u_0^\ast$ enforces the constraint $\dot u_L - ik u_0 = 0$, eliminating the pure-gauge longitudinal mode $u_L$ and leaving a single physical odd-parity tensor degree of freedom:
\begin{equation}
L_{\rm phys,pr}^{(2)} = \frac{K_{\rm T}}{2} |\dot u_T|^2 - \frac{F_{\rm T}}{2} k^2 |u_T|^2 ~.
\label{eq:s3}
\end{equation}
Varying $u_T^\ast$ gives the local high-frequency dispersion relation $-K_{\rm T}\omega^2 + F_{\rm T}k^2 = 0$ with $c_{\rm odd}^{\,2} = F_{\rm T}/K_{\rm T}$, as quoted in Eq.~\eqref{eq:oddchar}. Combining this with Eq.~\eqref{eq:s1-undivided} yields the turning-point identity~\eqref{eq:turn}, $(\ddot q/q)|_{\dot q=0} = c_{\rm odd}^{\,2}/b^2$. If $K_{\rm T} = 0$, the kinetic term in Eq.~\eqref{eq:s2} vanishes (so the odd-tensor principal symbol is already degenerate) and the undivided relation~\eqref{eq:s1-undivided} simultaneously forces $F_{\rm T} = 0$.

\section{General Quadratic HOST Theories and Tensor Characteristics}\label{sec:s2}

\subsection{Off-shell radial shape identity}
The general quadratic HOST action in Eq.~\eqref{eq:quad} is constructed from $\phi_{\mu\nu} \equiv \nabla_\mu\nabla_\nu\phi$ and the five quadratic contractions
\begin{equation}
\begin{aligned}
L_1 &= \phi_{\mu\nu}\phi^{\mu\nu} ~, &
L_2 &= (\Box\phi)^2 ~, \\
L_3 &= (\Box\phi)\,\phi^\mu\phi_{\mu\nu}\phi^\nu ~, &
L_4 &= \phi^\mu\phi_{\mu\rho}\phi^{\rho\nu}\phi_\nu ~, \\
L_5 &= (\phi^\mu\phi_{\mu\nu}\phi^\nu)^2 ~,
\end{aligned}
\label{eq:s4}
\end{equation}
with arbitrary coefficient functions $F_0, Q, F_2, A_{1,\dots,5}(\phi, X)$. We impose no degeneracy relation among these functions, so the derivation below applies simultaneously to nondegenerate quadratic HOST, quadratic GLPV, all quadratic DHOST subclasses, and quadratic U-DHOST.

To derive Eq.~\eqref{eq:quadidentity} without assuming a timelike scalar clock or a monotonic sphere radius, we work directly in the general static radial coordinates of Eq.~\eqref{eq:s0-metric} (covering both the static exterior $A, B > 0$ and the trapped interior $A, B < 0$). Define $\mathcal N \equiv \sqrt{|A|}$, $n \equiv \mathcal N'/\mathcal N = A'/(2A)$, $j \equiv J'/J$, $s \equiv \sqrt{A/B} > 0$, and $U \equiv A/J^2$, where primes denote $\dd/\dd x$. Integrating over the two-sphere, the four-dimensional volume element is proportional to $\sqrt{-g} \propto \mathcal N J^2/\sqrt{|B|} = s J^2$. We perform a volume-preserving shape variation by varying $\mathcal N$ and $J$ according to
\begin{equation}
\delta\mathcal N = 2\epsilon(x)\mathcal N ~, \qquad \delta J = -\epsilon(x) J ~, \qquad \delta B = 0 ~, \qquad \delta\phi = 0 ~,
\label{eq:s2-shape-var}
\end{equation}
which leaves the reduced volume measure $s J^2$, the kinetic invariant $X = B\phi'^2$, and all coefficient functions of $(\phi,X)$ strictly invariant.

Writing $w \equiv \phi'(x)$, the mixed Hessian $H^\mu{}_\nu \equiv \nabla^\mu\nabla_\nu\phi$ on the background~\eqref{eq:s0-metric} is diagonal with entries $(\lambda_t, \lambda_x, \lambda_\Omega, \lambda_\Omega)$ given by
\begin{equation}
\lambda_t = B n w ~, \qquad \lambda_x = B w' + \tfrac{1}{2} B' w ~, \qquad \lambda_\Omega = B j w ~.
\label{eq:hessian-eigenvalues}
\end{equation}
In terms of these eigenvalues, the scalar d'Alembertian and the five quadratic contractions~\eqref{eq:s4} read
\begin{equation}
\begin{gathered}
\Box\phi = \lambda_t + \lambda_x + 2\lambda_\Omega ~, \qquad
L_1 = \lambda_t^2 + \lambda_x^2 + 2\lambda_\Omega^2 ~, \\
L_3 = X\lambda_x\Box\phi ~, \qquad L_4 = X\lambda_x^2 ~, \qquad L_5 = X^2\lambda_x^2 ~.
\end{gathered}
\label{eq:hessian-contractions}
\end{equation}
Under the shape variation~\eqref{eq:s2-shape-var}, one has $\delta n = 2\epsilon'$ and $\delta j = -\epsilon'$, which implies
\begin{equation}
\delta\lambda_t = 2 B w \epsilon' ~, \qquad \delta\lambda_\Omega = -B w \epsilon' ~, \qquad \delta\lambda_x = 0 ~.
\end{equation}
Crucially, the trace variation vanishes identically, $\delta(\Box\phi) = \delta\lambda_t + 2\delta\lambda_\Omega = 0$, and since $\delta\lambda_x = 0$ and $\delta X = 0$, the variations of $L_2, L_3, L_4$, and $L_5$ all vanish: $\delta L_2 = \delta L_3 = \delta L_4 = \delta L_5 = 0$. The only nonvanishing variations come from $L_1$ and the Ricci scalar $R$:
\begin{equation}
\delta L_1 = 4 B X (n-j)\epsilon', \qquad
\delta R = \frac{4\epsilon}{J^2} - 4 B (n-j)\epsilon'.
\label{eq:shape-cancellations}
\end{equation}
The $\epsilon''$ terms cancel identically under the volume-preserving
shape variation, so the displayed expression for $\delta R$ is exact. Consequently, the operators $F_0, Q\Box\phi$, and $A_2 L_2, \dots, A_5 L_5$ drop out of the shape variation of the reduced Lagrangian $L_{\rm red} = s J^2 \mathcal L_{\rm quad}$, leaving only $F_2 R$ and $A_1 L_1$:
\begin{equation}
\delta L_{\rm red} = 4 s F_2 \epsilon - 4 s J^2 B (F_2 - X A_1)(n - j)\epsilon' ~.
\label{eq:s5}
\end{equation}
Noting that $U'/U = A'/A - 2J'/J = 2(n - j)$ and $B U = A U / s^2 = J^2 U^2 / s^2$, we have $2 s J^2 B (n - j) = J^4 U' / s$. Integrating Eq.~\eqref{eq:s5} by parts with $K_{\rm T} \equiv F_2 - X A_1$ yields the off-shell Euler identity
\begin{equation}
2\mathcal N E_{\mathcal N} - J E_J = 2\left[\frac{J^4 K_{\rm T}}{s}\,U'\right]' + 4 s F_2 ~,
\label{eq:s6}
\end{equation}
where $E_{\mathcal N}$ and $E_J$ are the Euler--Lagrange derivatives of the reduced action with respect to $\mathcal N$ and $J$. Setting $E_{\mathcal N} = E_J = 0$ on shell immediately gives Eq.~\eqref{eq:quadidentity}. Inside the $T$-region ($A, B < 0$), substituting $U = -q^2$ and $\dd/\dd x = (-B)^{-1/2}\dd/\dd\tau$ at a turning point $\dot q = 0$ ($U' = 0$) gives $U'' = 2 q\ddot q / B$, so Eq.~\eqref{eq:quadidentity} reduces directly to the undivided turning-point relation $K_{\rm T}(\ddot q/q)|_{\dot q=0} = F_2/J^2$.

It is worth emphasizing why this obstruction is not a Horndeski-type no-go that DHOST operators can circumvent. In spatially flat FRW cosmology, the Horndeski no-go for nonsingular bounces is evaded precisely because the new DHOST operators modify the \emph{scalar} (even-parity) dispersion relation, while the tensor sector is left untouched \cite{Ilyas:2020qja}. Here the logic is different: Eq.~\eqref{eq:s6} involves only $F_2$ and $K_{\rm T} = F_2 - XA_1$, which are exactly the two functions that fix the odd-parity principal symbol~\eqref{eq:s7}. A DHOST operator that shifts the tensor speed $c_{\rm odd}^{\,2} = F_2/K_{\rm T}$ therefore shifts the coefficient multiplying $\ddot q$ in Eq.~\eqref{eq:s1-undivided} by the same amount, so the product $K_{\rm T}(\ddot q/q)|_{\dot q=0}$ stays pinned to $F_2/J^2$ for arbitrary $F_2$ and $A_1$. The identity is thus a relation between the geometry and the tensor cone itself, rather than a statement about a particular operator truncation.

\subsection{Odd-parity tensor principal symbol and quadratic U-DHOST example}
For odd-parity tensor perturbations around the static background~\eqref{eq:s0-metric}, $\delta\phi = \delta X = 0$ and the linearized variations of $\Box\phi$ and $\phi^\mu\phi_{\mu\nu}\phi^\nu$ vanish by parity. Thus only $F_2 R$ and $A_1 L_1$ contribute to the two-derivative odd-parity action, which is governed by the effective principal metric $F_2 g^{\mu\nu} - A_1 \phi^\mu\phi^\nu$ \cite{Takahashi:2019oxz}. In local orthonormal coordinates (with frequency $\omega$, radial/longitudinal wave number $k_\parallel$ along $x$, and angular wave number $k_\perp$ along $S^2$), the principal symbol takes the form
\begin{equation}
P_{\rm odd}(\omega, k_\parallel, k_\perp) =
\begin{cases}
-F_2\omega^2 + K_{\rm T}k_\parallel^2 + F_2 k_\perp^2 ~, & A, B > 0 \quad (\text{static exterior}), \\[4pt]
-K_{\rm T}\omega^2 + F_2\left(k_\parallel^2 + k_\perp^2\right) ~, & A, B < 0 \quad (\text{trapped interior}).
\end{cases}
\label{eq:s7}
\end{equation}
Notice how the exchange of the timelike and spatial coordinates across the horizon swaps the roles of $F_2$ and $K_{\rm T}$: inside the $T$-region ($A, B < 0$), the odd-parity dispersion relation is isotropic with $c_{\rm odd}^{\,2} = F_2/K_{\rm T}$, whereas in the static exterior ($A, B > 0$), radial propagation ($k_\perp = 0$) has squared speed $c_{r,\rm T}^{\,2} = K_{\rm T}/F_2$. Consequently, $F_2/K_{\rm T} < 0$ inside the $T$-region makes the principal symbol elliptic (definite signature), while $K_{\rm T}/F_2 < 0$ in the exterior gives the principal metric split signature $(2,2)$.

\paragraph{Explicit quadratic U-DHOST example.}
The quadratic U-DHOST family has the exact action~\eqref{eq:quad}, with its coefficients constrained so that the theory is degenerate in unitary gauge ($X < 0$) while propagating a harmless ``shadowy'' mode \cite{DeFelice:2018ewo,DeFelice:2021hps}. Because Eqs.~\eqref{eq:s5}--\eqref{eq:s7} hold for arbitrary $F_2$ and $A_{1,\dots,5}$ prior to imposing any degeneracy relation, quadratic U-DHOST theories automatically belong to TPST. As a concrete illustration, consider the U-DHOST model of Eq.~(6) in Ref.~\cite{DeFelice:2021hps}:
\begin{equation}
F_2 = \frac{M_{\rm Pl}^2}{2} ~, \qquad A_1 = A_2 = A_3 = 0 ~, \qquad A_4 = -\xi X ~, \qquad A_5 = \xi ~.
\label{eq:udhost-example}
\end{equation}
Because $A_4 L_4$ and $A_5 L_5$ neither vary under the volume-preserving shape transformation ($\delta L_4 = \delta L_5 = 0$) nor contribute to the odd-parity tensor sector, this U-DHOST deformation leaves $K_{\rm T} = F_{\rm T} = M_{\rm Pl}^2/2$ and $(\ddot q/q)|_{\dot q=0} = 1/b^2 > 0$ untouched for all values of $\xi$, ruling out any two-horizon RBH interior. (Note that inside the $T$-region, a nonvanishing radial scalar gradient satisfies $X = B\phi'^2 < 0$, which is precisely the timelike-gradient regime where the unitary-gauge shadowy condition is defined.)

\section{Full Horndeski Theory and the Quintic Interaction}\label{sec:s3}
Horndeski theory up to quartic order ($G_2, G_3, G_4$) is already contained in the quadratic HOST action~\eqref{eq:quad} with $F_0 = G_2$, $Q = -G_3$, $F_2 = G_4$, $A_1 = -A_2 = -G_{4,\chi}$ (where $\chi \equiv -X/2$), and $A_3 = A_4 = A_5 = 0$. The quintic Horndeski interaction $G_5(\phi,\chi)$ requires a separate verification because it contains terms cubic in $H^\mu{}_\nu \equiv \nabla^\mu\nabla_\nu\phi$. In standard notation~\cite{Minamitsuji:2022vbi} with $[H^n] \equiv \operatorname{tr}(H^n)$, the full Horndeski Lagrangian reads
\begin{equation}
\mathcal L_{\rm H} = G_2 - G_3[H] + G_4 R + G_{4\chi}\bigl([H]^2 - [H^2]\bigr)
+ G_5 G^\mu{}_\nu H^\nu{}_\mu - \frac{G_{5\chi}}{6}\bigl([H]^3 - 3[H][H^2] + 2[H^3]\bigr) ~.
\label{eq:horndeski-action}
\end{equation}

\subsection{Off-shell shape identity for full Horndeski theory}
On the general static metric~\eqref{eq:s0-metric}, the eigenvalues of $H^\mu{}_\nu$ are given by Eq.~\eqref{eq:hessian-eigenvalues}, while the mixed components of the Ricci tensor $R^\mu{}_\nu$ are diagonal with eigenvalues
\begin{equation}
\begin{aligned}
R^t{}_t &= -B(n' + n^2) - \left(\frac{B'}{2} + 2 B j\right)n ~, \\
R^x{}_x &= -B(n' + n^2) - \frac{B' n}{2} - 2 B(j' + j^2) - B' j ~, \\
R^\theta{}_\theta = R^\varphi{}_\varphi &= \frac{1}{J^2} - B(j' + 2 j^2) - \left(B n + \frac{B'}{2}\right)j ~.
\end{aligned}
\label{eq:ricci-eigenvalues}
\end{equation}
Substituting Eqs.~\eqref{eq:hessian-eigenvalues} and \eqref{eq:ricci-eigenvalues} into the reduced Lagrangian $L_{\rm red} = s J^2 \mathcal L_{\rm H}$ and forming the volume-preserving Euler combination $2\mathcal N E_{\mathcal N} - J E_J$ (corresponding to $\delta\mathcal N = 2\epsilon\mathcal N$, $\delta J = -\epsilon J$ at fixed $B, \phi$), the $G_2$ and $G_3$ contributions vanish identically because $\delta\chi = \delta[H] = 0$. The $G_4$ and $G_5$ contributions organize into total radial derivatives plus curvature terms:
\begin{equation}
2\mathcal N E_{\mathcal N}^{(i)} - J E_J^{(i)} = \left[\frac{J^4 \mathcal H_i}{s}\,U'\right]' + 2 s \mathcal F_i ~,
\qquad i = 4, 5 ~,
\label{eq:horndeski-termwise}
\end{equation}
where explicit evaluation gives
\begin{equation}
\begin{aligned}
\mathcal F_4 &= 2 G_4 ~, &
\mathcal H_4 &= 2 G_4 + 2 B\phi'^2 G_{4\chi} ~, \\
\mathcal F_5 &= B\phi'^2 G_{5\phi} - \tfrac{1}{2} B B'\phi'^3 G_{5\chi} - B^2\phi'^2\phi'' G_{5\chi} ~, &
\mathcal H_5 &= -B\phi'^2 G_{5\phi} - B^2\frac{J'}{J}\phi'^3 G_{5\chi} ~.
\end{aligned}
\label{eq:horndeski-pieces}
\end{equation}
Combining $\mathcal F \equiv \mathcal F_4 + \mathcal F_5$ and $\mathcal H \equiv \mathcal H_4 + \mathcal H_5$ yields the full Horndeski off-shell shape identity:
\begin{equation}
2\mathcal N E_{\mathcal N} - J E_J = \left[\frac{J^4 \mathcal H}{s}\,U'\right]' + 2 s \mathcal F ~.
\label{eq:horndeski-offshell}
\end{equation}
On shell ($E_{\mathcal N} = E_J = 0$), Eq.~\eqref{eq:horndeski-offshell} has the exact form of Eq.~\eqref{eq:quadidentity} with $(K_{\rm T}, F_2) \to (\mathcal H/2, \mathcal F/2)$, and in areal coordinates ($J(r) = r$) it reproduces the Horndeski specialization of Eq.~(80) of Ref.~\cite{Boonserm:2026qwz}. Just as in Appendix~\ref{sec:s2}, Eq.~\eqref{eq:horndeski-offshell} involves no division by $J'$ or $\phi'$.

\subsection{Coincidence of odd-parity tensor coefficients at the turning point}
In full Horndeski theory, the quadratic action for odd-parity perturbations involves three coefficients $\mathcal F, \mathcal G, \mathcal H$~\cite{Minamitsuji:2022vbi}, where $\mathcal F$ and $\mathcal H$ are given by Eq.~\eqref{eq:horndeski-pieces} and
\begin{equation}
\mathcal G = 2 G_4 + B\phi'^2(2 G_{4\chi} - G_{5\phi}) - \frac{B^2 A'}{2A}\phi'^3 G_{5\chi}
\quad\Longrightarrow\quad
\mathcal G - \mathcal H = -\frac{B^2\phi'^3 G_{5\chi}}{2}\left(\frac{A'}{A} - \frac{2J'}{J}\right) = -\frac{B^2\phi'^3 G_{5\chi}}{2}\frac{U'}{U} ~.
\label{eq:horndeski-directional}
\end{equation}
Away from the turning point, $G_{5\chi} \neq 0$ splits $\mathcal G$ and $\mathcal H$, making the longitudinal and angular odd-tensor propagation speeds distinct. At the turning point $U' = 0$ ($\dot q = 0$), however, this splitting vanishes identically ($\mathcal G = \mathcal H$), so the odd-parity principal symbol becomes isotropic, $-\mathcal H\omega^2 + \mathcal F|\boldsymbol k|^2 = 0$, with $c_{\rm odd}^{\,2} = \mathcal F/\mathcal H$. Converting Eq.~\eqref{eq:horndeski-offshell} to proper time $\tau$ inside the $T$-region gives
\begin{equation}
\frac{\dd}{\dd\tau}\!\left(v\mathcal H\frac{\dot q}{q}\right) = a\mathcal F
\quad\Longrightarrow\quad
\left.\mathcal H\frac{\ddot q}{q}\right|_{\dot q=0} = \frac{\mathcal F}{b^2} ~,
\label{eq:horndeski-turn}
\end{equation}
which reproduces the TPST turning-point identity~\eqref{eq:turn} when $\mathcal H \neq 0$ (and forces $\mathcal F = 0$ if $\mathcal H = 0$).

\section{Disformal Invariance and Higher-Derivative Field Transformations}\label{sec:s4}

\subsection{General proof of disformal invariance near the turning point}
Consider a theory with action $S[g,\phi] = S_{\rm seed}[\bar g,\bar\phi]$ obtained from a TPST seed theory $S_{\rm seed}$ via a field transformation $T\colon (g_{\mu\nu},\phi) \mapsto (\bar g_{\mu\nu},\phi)$. We assume that in an open neighborhood of the interior turning point of $q$:
(i) $T$ is smooth and possesses a smooth local functional inverse in field space;
(ii) on the homogeneous $T$-region background, $T$ rescales the spatial scale factors isotropically ($\bar a = \sqrt{C}\,a$, $\bar b = \sqrt{C}\,b$) and the proper time via $\dd\bar\tau = \sqrt{E}\,\dd\tau$ with finite $C, E > 0$; and
(iii) the odd-parity perturbation sector remains closed under the linearized map $\mathcal J \equiv DT$, whose restriction to the odd sector is a nonsingular algebraic matrix $M$ ($\delta\bar g_{\rm odd} = M\,\delta g_{\rm odd}$) containing no derivatives of the perturbations.

Condition~(i) guarantees that any background solution $(g,\phi)$ of $S$ corresponds to a background solution $(\bar g,\phi)$ of $S_{\rm seed}$ ($E_{\rm seed}[\bar g,\phi] = 0$). Expanding $S = S_{\rm seed} \circ T$ to second order around a solution therefore eliminates the term proportional to $E_{\rm seed}$:
\begin{equation}
\delta^2 S = \mathcal J^\dagger\,\delta^2 S_{\rm seed}\,\mathcal J + E_{\rm seed}\mathbin{\cdot}\delta^2 T
= \mathcal J^\dagger\,\delta^2 S_{\rm seed}\,\mathcal J ~.
\label{eq:action-hessian}
\end{equation}
By condition~(iii), restricting Eq.~\eqref{eq:action-hessian} to the physical odd-parity sector at high frequency relates the principal symbols by the algebraic congruence $P_{\rm odd}(k) = M^{\mathrm T}\bar P_{\rm odd}(k) M$ of Eq.~\eqref{eq:congruence} (where derivatives of the background-dependent matrix $M$ contribute only subleading lower-derivative terms). By Sylvester's law of inertia, congruence under a real nonsingular matrix $M$ preserves both the rank (nondegeneracy) and the signature (hyperbolicity) of the odd-parity principal symbol. Meanwhile, condition~(ii) implies $\bar q = \bar a/\bar b = a/b = q$ and $(\dd^2\bar q/\dd\bar\tau^2)|_{\dot q=0} = \ddot q_\ast / E$ (Eq.~\eqref{eq:transportshape}). In local orthonormal frames, frequencies and wave numbers rescale as $\bar\omega = \omega/\sqrt{E}$ and $\bar k = k/\sqrt{C}$, so $c_{\rm odd}^{\,2} = (E/C)\bar c_{\rm odd}^{\,2}$ and the seed turning-point identity transforms as
\begin{equation}
\frac{\ddot q_\ast}{q_\ast} = \frac{E\,\bar c_{\rm odd}^{\,2}}{\bar b_\ast^2} = \frac{c_{\rm odd}^{\,2}}{b_\ast^2} ~.
\label{eq:transport-turn}
\end{equation}
Notice that conditions~(i)--(iii) are required only in an open neighborhood of the turning point $\tau_\ast$; the transformation $T$ need not remain regular at the horizons $x_\pm$.

\subsection{Ordinary disformal maps and cubic GLPV/DHOST theories}\label{sec:s4b}
For a standard disformal transformation $\bar g_{\mu\nu} = C(\phi,X)g_{\mu\nu} + D(\phi,X)\phi_\mu\phi_\nu$ with $\bar\phi = \phi$, the transformed kinetic invariant and Jacobian are
\begin{equation}
\bar X = \frac{X}{E} ~, \qquad
\frac{\partial\bar X}{\partial X} = \frac{C - X C_X - X^2 D_X}{E^2} ~, \qquad E \equiv C + D X ~.
\label{eq:ordinary-disformal}
\end{equation}
Whenever $C > 0$, $E > 0$, and $\partial\bar X/\partial X \neq 0$ near the turning point, the implicit function theorem guarantees that $X$ can be solved locally in terms of $(\phi,\bar X)$, yielding the explicit inverse $g_{\mu\nu} = C^{-1}(\bar g_{\mu\nu} - D\phi_\mu\phi_\nu)$. On the homogeneous interior background, $\bar a = \sqrt{C}\,a$, $\bar b = \sqrt{C}\,b$, and $\dd\bar\tau = \sqrt{E}\,\dd\tau$. In the odd-parity perturbation sector, $\delta\phi = \delta X = 0$, so $\delta\bar g_{\rm odd} = C\,\delta g_{\rm odd}$ is purely algebraic and nonsingular. All three conditions~(i)--(iii) are therefore satisfied, extending our no-go theorem to all cubic GLPV and DHOST theories in the regular disformal orbit of Horndeski theory~\cite{BenAchour:2016fzp}.

\subsection{Higher-derivative scalar-invariant-gradient maps}\label{sec:s4c}
More generally, consider a metric transformation built from finite-order, parity-even scalar differential invariants $I_A(g,\phi)$:
\begin{equation}
\bar g_{\mu\nu} = C(I)\,g_{\mu\nu} + \sum_{A,B} D_{AB}(I)\,\partial_\mu I_A\,\partial_\nu I_B ~, \qquad \bar\phi = \phi ~.
\label{eq:s10}
\end{equation}
On the homogeneous interior background, every scalar invariant depends only on $\tau$, $I_A = I_A(\tau)$, so the gradient terms $\partial_\mu I_A\,\partial_\nu I_B$ modify only the $\dd\tau^2$ component with $E = C - \sum_{A,B} D_{AB}\dot I_A\dot I_B$. For odd-parity perturbations with $\ell \ge 2$, rotational parity forbids any linear variation of a parity-even scalar invariant ($\delta I_A = 0$), which immediately gives $\delta\bar g_{\rm odd} = C\,\delta g_{\rm odd}$. Thus, whenever Eq.~\eqref{eq:s10} admits a smooth local functional inverse and satisfies $C, E > 0$ near the turning point, the TPST obstruction is preserved.

As an explicit higher-derivative example on $X \neq 0$, define the projected derivative $\mathrm D_\mu\psi \equiv (\delta_\mu^\nu - \phi_\mu\phi^\nu/X)\partial_\nu\psi$ and $Z_\perp \equiv \mathrm D_\mu X\,\mathrm D^\mu X$. For any constant $\lambda$, the generalized-disformal map~\cite{Takahashi:2021ttd,Takahashi:2022mew}
\begin{equation}
T_\lambda(g)_{\mu\nu} = g_{\mu\nu} + \lambda^2 Z_\perp\phi_\mu\phi_\nu + 2\lambda\phi_{(\mu}\mathrm D_{\nu)}X
\label{eq:s11}
\end{equation}
satisfies
\begin{equation}
\bar X = X ~, \qquad \bar Z_\perp = Z_\perp ~, \qquad \bar{\mathrm D}_\mu X = \mathrm D_\mu X + \lambda Z_\perp\phi_\mu ~, \qquad
T_{-\lambda} \circ T_\lambda = \mathrm{id} ~, \qquad \frac{\det\bar g}{\det g} = 1 ~,
\label{eq:s12}
\end{equation}
with exact inverse metric
\begin{equation}
\bar g^{\mu\nu} = g^{\mu\nu} - \lambda(\phi^\mu u^\nu + u^\mu\phi^\nu) + \lambda^2 X u^\mu u^\nu ~, \qquad u_\mu \equiv \mathrm D_\mu X ~.
\label{eq:s13}
\end{equation}
Because $T_{-\lambda}$ provides an exact functional inverse on an open neighborhood in field space and $u_\mu = 0$ on the radial background (giving $C = E = 1$ and $\delta\bar g_{\rm odd} = \delta g_{\rm odd}$), pulling back any TPST seed action under $T_\lambda$ yields a generalized-disformal theory that inherits the no-go theorem~\cite{Takahashi:2021ttd,Takahashi:2023vva}.

\subsection{Invertible Weyl-curvature conformal maps}\label{sec:s4d}
To illustrate a transformation depending directly on the Riemann curvature, let $W \equiv C_{\alpha\beta\gamma\delta}C^{\alpha\beta\gamma\delta}$ denote the Weyl scalar and consider the conformal transformation $\bar g_{\mu\nu} = \Omega(W)\,g_{\mu\nu}$ in four dimensions. Because $\bar W = W / \Omega(W)^2$, $W$ can be uniquely inverted as a smooth function of $\bar W$ whenever
\begin{equation}
\Omega(W) > 0 ~, \qquad \Omega(W) - 2 W \Omega_W(W) \neq 0 ~,
\label{eq:weyl-map}
\end{equation}
which occurs, for example, for $\Omega(W) = 1 + \lambda W$ on $|\lambda W| < 1$. Because $W$ is a parity-even scalar invariant, its linear odd-parity variation around a spherically symmetric background vanishes ($\delta W = 0$), so $\delta\bar g_{\rm odd} = \Omega(W)\,\delta g_{\rm odd}$ is algebraic and nonsingular. Pulling back the Einstein--Hilbert (or $k$-essence) action under $\bar g_{\mu\nu} = \Omega(W)\,g_{\mu\nu}$ yields, up to a boundary term,
\begin{equation}
S = F_\star\int\dd^4x\sqrt{-g}\left[\Omega(W) R + \frac{3}{2\Omega(W)}\nabla_\mu\Omega(W)\nabla^\mu\Omega(W)\right] + S_{\text{matter}}[\Omega(W)g,\phi] ~,
\label{eq:weyl-action}
\end{equation}
which contains higher derivatives of the Weyl tensor yet remains subject to the turning-point no-go theorem on the invertible domain~\eqref{eq:weyl-map}.

\section{Summary of Theory Coverage and Direct Extensions}\label{sec:s5}
Table~\ref{tab:coverage} summarizes the classes of scalar--tensor and modified gravity theories covered by our interior turning-point obstruction. Beyond the families connected to TPST seeds by invertible transformations, the obstruction also extends directly to two classes of modifications that leave both the homogeneous shape variation and the quadratic odd-parity principal symbol unchanged:

\begin{table}[htbp]
\centering
\caption{Summary of gravitational theory families covered by the interior turning-point no-go theorem.}
\label{tab:coverage}
\small\renewcommand{\arraystretch}{1.18}
\begin{tabular}{@{}l@{\hspace{0.02\linewidth}}l@{\hspace{0.02\linewidth}}l@{}}
\toprule
\parbox[t]{0.23\linewidth}{\raggedright \textbf{Theory family}} & \parbox[t]{0.50\linewidth}{\raggedright \textbf{Included domain}} & \parbox[t]{0.21\linewidth}{\raggedright \textbf{Derivation}} \\
\midrule
\parbox[t]{0.23\linewidth}{\raggedright $k$-essence \& SCG foliation} & \parbox[t]{0.50\linewidth}{\raggedright $k$-essence~\eqref{eq:kessence} and the general ADM action~\eqref{eq:foliation} with a timelike interior scalar gradient ($\dot\phi\neq 0$)} & \parbox[t]{0.21\linewidth}{\raggedright Appendix~\ref{sec:s1}} \\
\parbox[t]{0.23\linewidth}{\raggedright General quadratic HOST} & \parbox[t]{0.50\linewidth}{\raggedright Arbitrary $F_0,Q,F_2,A_{1,\dots,5}(\phi,X)$ in Eq.~\eqref{eq:quad}, including quadratic GLPV, DHOST (classes Ia/IIa), and U-DHOST} & \parbox[t]{0.21\linewidth}{\raggedright Appendix~\ref{sec:s2}} \\
\parbox[t]{0.23\linewidth}{\raggedright Full Horndeski theory} & \parbox[t]{0.50\linewidth}{\raggedright Arbitrary $G_2,G_3,G_4,G_5(\phi,\chi)$ in Eq.~\eqref{eq:horndeski-action}} & \parbox[t]{0.21\linewidth}{\raggedright Appendix~\ref{sec:s3}} \\
\parbox[t]{0.23\linewidth}{\raggedright Invertible disformal orbit} & \parbox[t]{0.50\linewidth}{\raggedright Cubic GLPV and DHOST images of Horndeski satisfying $C,E>0$ and $\partial\bar X/\partial X\neq 0$ near the turning point} & \parbox[t]{0.21\linewidth}{\raggedright Appendix~\ref{sec:s4b}} \\
\parbox[t]{0.23\linewidth}{\raggedright Higher-derivative images} & \parbox[t]{0.50\linewidth}{\raggedright Admissible scalar-invariant-gradient maps~\eqref{eq:s10} (e.g.\ $T_\lambda$ in Eq.~\eqref{eq:s11}) and Weyl-conformal maps~\eqref{eq:weyl-map}} & \parbox[t]{0.21\linewidth}{\raggedright Appendices~\ref{sec:s4c}--\ref{sec:s4d}} \\
\parbox[t]{0.23\linewidth}{\raggedright Higher spatial derivatives} & \parbox[t]{0.50\linewidth}{\raggedright SCG operators with vanishing homogeneous shape and odd principal variations, e.g.\ $f(\eta,N)D_i\mathcal R D^i\mathcal R$} & \parbox[t]{0.21\linewidth}{\raggedright Eq.~\eqref{eq:invisible-operator}} \\
\parbox[t]{0.23\linewidth}{\raggedright Mimetic gravity} & \parbox[t]{0.50\linewidth}{\raggedright Any TPST seed action supplemented by the Lagrange multiplier constraint $\lambda[X - X_\star(\phi)]$} & \parbox[t]{0.21\linewidth}{\raggedright Appendix~\ref{sec:s5}, item~(ii)} \\
\bottomrule
\end{tabular}
\end{table}

\paragraph{(i) Higher-spatial-derivative SCG operators.}
Adding any operator to the SCG action~\eqref{eq:foliation} whose variation with respect to $\beta$ on the homogeneous interior vanishes and whose quadratic odd-parity variation also vanishes leaves Eq.~\eqref{eq:turn} intact. For example, consider
\begin{equation}
\Delta S = \int\dd\eta\,\dd^3x\,N\sqrt{h}\,f(\eta,N)\,D_i\mathcal R\,D^i\mathcal R ~.
\label{eq:invisible-operator}
\end{equation}
Because the spatial Ricci scalar $\mathcal R = 2/b(\eta)^2$ is homogeneous on constant-$\eta$ slices ($D_i\mathcal R = 0$) and its linear odd-parity variation vanishes ($\delta\mathcal R = 0$), $\Delta S$ starts at second order in spatial gradients of even-parity perturbations and contributes nothing to either the background shape equation or the quadratic odd-parity action.

\paragraph{(ii) Noninvertible mimetic constraints.}
In mimetic gravity~\cite{Arroja:2015wpa}, a seed action $S_{\rm seed}[g,\phi]$ is supplemented by a Lagrange multiplier enforcing the constraint $X = X_\star(\phi)$:
\begin{equation}
S_{\rm mim} = S_{\rm seed}[g,\phi] + \int\dd^4x\sqrt{-g}\,\lambda\bigl[X - X_\star(\phi)\bigr] ~.
\label{eq:s5-mimetic}
\end{equation}
Although the corresponding disformal parametrization of the metric is noninvertible (so Appendix~\ref{sec:s4} does not apply), the coverage of Eq.~\eqref{eq:s5-mimetic} follows directly from the volume-preserving shape variation~\eqref{eq:s2-shape-var}. Because $\delta\sqrt{-g} = 0$ and $\delta X = 0$ under Eq.~\eqref{eq:s2-shape-var}, the constraint term $\sqrt{-g}\,\lambda[X - X_\star(\phi)]$ has identically zero shape variation, and because $\delta\phi = \delta\lambda = \delta X = 0$ in the odd-parity sector, it introduces no two-derivative terms into the odd-parity principal symbol. Thus any TPST seed theory supplemented by the mimetic constraint ~\eqref{eq:s5-mimetic} obeys the exact same turning-point identity~\eqref{eq:turn}.

\section{Extremal Exterior Corollary and Explicit Class-Ia DHOST Examples}\label{sec:s7}

\subsection{Double-zero extremal horizon and finite-order coefficient zeros}
We now prove the extremal exterior corollary, Eq.~\eqref{eq:extremal}, using the quadratic HOST radial identity~\eqref{eq:quadidentity} on the static exterior ($r > r_e$) alone. In areal coordinates ($J(r) = r$), let $\rho \equiv r - r_e > 0$ denote the radial distance outside an extremal outer horizon at $r_e > 0$, where $A(r)$ and $B(r)$ have a double zero:
\begin{equation}
A(r) = a_2\rho^2 + \mathcal O(\rho^3) ~, \qquad B(r) = b_2\rho^2 + \mathcal O(\rho^3) ~, \qquad a_2, b_2 > 0 ~,
\label{eq:extremal-expansion}
\end{equation}
corresponding to a near-horizon $\mathrm{AdS}_2 \times S^2$ geometry with $\mathrm{AdS}_2$ radius $L_2 = b_2^{-1/2}$. Defining $\mathcal P(r) \equiv r^4 K_{\rm T}/s$ with $s = \sqrt{A/B}$ and $U = A/r^2$, assume that $A/\rho^2, B/\rho^2, K_{\rm T}$, and $F_2$ extend smoothly to $\rho = 0^+$ with finite nonzero $K_{{\rm T},e}$ and $F_{2,e}$. Then $s \to s_e = \sqrt{a_2/b_2}$ and $U'(\rho) = (2a_2/r_e^2)\rho + \mathcal O(\rho^2)$, so $\mathcal P' U' = \mathcal O(\rho) \to 0$ as $\rho \to 0^+$. Taking the $\rho \to 0^+$ limit of Eq.~\eqref{eq:quadidentity}, $(\mathcal P U')' = -2s F_2$, yields
\begin{equation}
\lim_{\rho\to 0^+}(\mathcal P U')' = \frac{2 a_2 r_e^2 K_{{\rm T},e}}{s_e} = -2 s_e F_{2,e}
\quad\Longrightarrow\quad
F_{2,e} = -r_e^2 b_2 K_{{\rm T},e} ~.
\label{eq:extremal-undivided}
\end{equation}
Dividing by $F_{2,e} \neq 0$ and identifying the radial exterior odd-tensor speed squared $c_{r,\rm T}^{\,2} = K_{\rm T}/F_2$ from Eq.~\eqref{eq:s7} gives Eq.~\eqref{eq:extremal}:
\begin{equation}
c_{r,\rm T,e}^{\,2} = \frac{K_{{\rm T},e}}{F_{2,e}} = -\frac{1}{r_e^2 b_2} = -\frac{L_2^2}{r_e^2} < 0 ~.
\label{eq:extremal-ratio}
\end{equation}

More generally, even if $K_{\rm T}$ vanishes at the extremal horizon to finite order $p \ge 1$,
\begin{equation}
K_{\rm T}(\rho) = k_p\rho^p + \mathcal O(\rho^{p+1}) ~, \qquad k_p \neq 0 ~, \quad p \in \{0, 1, 2, \dots\} ~,
\label{eq:finite-order-kinetic}
\end{equation}
substituting Eq.~\eqref{eq:finite-order-kinetic} into $(\mathcal P U')' = -2s F_2$ fixes the leading behavior of $F_2(\rho)$ as $\rho \to 0^+$:
\begin{equation}
F_2(\rho) = -r_e^2 b_2 (p + 1) k_p \rho^p + \mathcal O(\rho^{p+1})
\quad\Longrightarrow\quad
\lim_{\rho\to 0^+} c_{r,\rm T}^{\,2}(\rho) = \lim_{\rho\to 0^+}\frac{K_{\rm T}}{F_2} = -\frac{L_2^2}{(p + 1)r_e^2} < 0 ~.
\label{eq:finite-order-ratio}
\end{equation}
Thus $K_{\rm T}$ and $F_2$ still have opposite signs on an open punctured exterior interval $\rho \in (0, \rho_0)$, where the odd-parity tensor principal symbol~\eqref{eq:s7} has split signature $(2,2)$. Furthermore, for any continuous one-parameter family of exterior solutions parameterized by mass $\mu$ (with extremal limit $\mu \to \mu_e^+$ and outer horizon $r_+(\mu) \to r_e^+$), joint continuity of $K_{\rm T}(r,\mu)$ and $F_2(r,\mu)$ guarantees that this opposite-sign instability band persists outside the outer horizon $r_+(\mu)$ for near-extremal black holes ($\mu$ sufficiently close to $\mu_e$).

\subsection{Class-Ia DHOST Hayward realization and the near-extremal mass window}
An explicit example is provided by the purely radial branch of the class-Ia DHOST Hayward geometry constructed in Secs.~2.1.1 and 3 of Ref.~\cite{Babichev:2020qpr}. Translating the notation of Ref.~\cite{Babichev:2020qpr} ($G, K, \mathcal H$) to ours gives $F_2 = G$, $F_0 = K$, $Q = 0$, and $K_{\rm T} = -\mathcal H$, with metric functions $h \equiv A$, $f \equiv B$ and kinetic invariant $X(r)$ displayed in Eq.~\eqref{eq:hayward} for constants $\alpha, \beta, \gamma > 0$. On the static exterior ($A = B > 0$), $X(r) > 0$ corresponds to a real radial scalar field ($\phi'^2 = X/B > 0$). Defining $S \equiv -H_0\beta^2 > 0$ and the dimensionless variable $y(r) \equiv \beta X(r)/\alpha = \beta r^3/(\beta r^3 + \gamma) \in (0,1)$, the odd-parity tensor coefficients read
\begin{equation}
\frac{K_{\rm T}}{S} = \frac{1}{y^2} > 0 ~, \qquad
\frac{F_2}{S} = \frac{6y - 5}{y^2} ~, \qquad
c_{r,\rm T}^{\,2} = \frac{K_{\rm T}}{F_2} = \frac{1}{6y - 5} ~,
\label{eq:s15}
\end{equation}
while the remaining quadratic DHOST couplings are $A_1 = -A_2 = 6 H_0\alpha(\alpha - \beta X)/X^3$, $A_3 = -4 F_{2X}/X + 2 A_1/X - 8 K_{\rm T}/(3 X^2)$, and $A_4, A_5$ fixed by the class-Ia degeneracy relations (since $X A_3 + 4 F_{2X} - 2 A_1 = -8 K_{\rm T}/(3X) \neq 0$, this model lies strictly outside the quadratic GLPV subclass).

To determine the near-extremal exterior exclusion window in Eq.~\eqref{eq:masswindow}, note that the horizon condition $A(r_+) = 0$ relates the mass parameter $\mu$ to the outer horizon radius $r_+$ via
\begin{equation}
\mu(r_+) = \frac{\beta r_+^3 + \gamma}{\alpha r_+^2} ~, \qquad
\frac{\dd\mu}{\dd r_+} = \frac{\beta - 2\gamma / r_+^3}{\alpha} ~.
\label{eq:s16}
\end{equation}
The minimum mass occurs at the extremal horizon radius $r_e = (2\gamma/\beta)^{1/3}$, where $\mu_e \equiv \mu(r_e) = 3\beta r_e/(2\alpha) = 3\beta^{2/3}\gamma^{1/3}/(2^{2/3}\alpha)$ and $y(r_e) = 2/3$, giving $c_{r,\rm T,e}^{\,2} = (4 - 5)^{-1} = -1$ in exact agreement with Eq.~\eqref{eq:extremal-ratio}. By Eq.~\eqref{eq:s15}, $F_2(r) < 0 < K_{\rm T}(r)$ whenever $y(r) < 5/6$, which corresponds to $r < r_0 \equiv (5\gamma/\beta)^{1/3} = (5/2)^{1/3} r_e$. Because $\mu(r_+)$ is strictly increasing for $r_+ > r_e$, the instability region $r_+ < r < r_0$ lies outside the outer horizon if and only if $r_e < r_+ < r_0$, or equivalently
\begin{equation}
1 < \frac{\mu}{\mu_e} < \frac{\mu(r_0)}{\mu_e} = \frac{2(R_0^3 + 1/2)}{3 R_0^2}\bigg|_{R_0=(5/2)^{1/3}} = 2\left(\frac{2}{5}\right)^{2/3} = 1.1 ~,
\end{equation}
which establishes Eq.~\eqref{eq:masswindow}.

\subsection{Independent check in the class-Ia DHOST arctangent family}
As a second, independent exterior check, consider the class-Ia DHOST arctangent family of Ref.~\cite{Baake:2021jzv}, whose metric functions $h \equiv A$, $f_p \equiv B$ and kinetic profile $X(r)$ are given by
\begin{equation}
A(r) = B(r) = 1 - \frac{2\mu X(r)}{r} ~, \qquad
X(r) = \frac{2}{\pi}\arctan\!\left(\frac{\pi r^p}{2\sigma^{p-1}}\right) ~, \qquad \sigma > 0 ~, \quad p \ge 3 ~.
\label{eq:arctan-profile}
\end{equation}
On the static exterior, the supporting class-Ia DHOST action has odd-parity tensor coefficients
\begin{equation}
\begin{gathered}
D_p(X) = p\sin(\pi X) - 3\pi X ~, \qquad
N_p(X) = p^2\sin(2\pi X) - 8p\sin(\pi X) + 6\pi X ~, \\
F_2 = \frac{N_p}{D_p^2} ~, \qquad K_{\rm T} = -\frac{2}{D_p} ~.
\end{gathered}
\label{eq:arctan-coefficients}
\end{equation}
At the extremal horizon $r = r_e$, the double-zero conditions $B(r_e) = B'(r_e) = 0$ require $r_e X'(r_e)/X(r_e) = 1$. Differentiating Eq.~\eqref{eq:arctan-profile} gives $r X'/X = p\sin(\pi X)/(\pi X)$, so $X_e \equiv X(r_e) \in (0,1)$ is the unique root of $p\sin(\pi X_e) = \pi X_e$. Substituting this root into Eq.~\eqref{eq:arctan-coefficients} and into $b_2 = \frac{1}{2}B''(r_e)$ yields $D_{p,e} = -2\pi X_e$ and $N_{p,e} = 2\pi X_e[p\cos(\pi X_e) - 1]$, from which we obtain
\begin{equation}
r_e^2 b_2 = \frac{1 - p\cos(\pi X_e)}{2} > 0 ~, \qquad
c_{r,\rm T,e}^{\,2} = \frac{K_{{\rm T},e}}{F_{2,e}} = \frac{2}{p\cos(\pi X_e) - 1} = -\frac{1}{r_e^2 b_2} < 0 ~.
\label{eq:arctan-check}
\end{equation}
Equation~\eqref{eq:arctan-check} agrees identically with the general geometric corollary~\eqref{eq:extremal-ratio} for all $p \ge 3$ (giving, e.g., $c_{r,\rm T,e}^{\,2} = -0.68$ for $p=3$ and $-0.48$ for $p=4$).

\bibliography{references}
\end{document}